\documentclass[sigconf]{acmart}
\usepackage{booktabs} 
\usepackage[T1]{fontenc}
\usepackage{graphicx}
\usepackage{amsmath}
\usepackage{esvect}
\usepackage{ulem}
\usepackage{xcolor}
\usepackage{colortbl}
\usepackage{multirow}
\usepackage{enumitem}

\newcommand*{\smallerstyle}[1]{%
   \ifx#1\displaystyle\scriptstyle\else
   \ifx#1\textstyle\scriptstyle\else
   \scriptscriptstyle\fi\fi
}

\begin{document}

\makeatletter
\newcommand*{\overmapsto}{\mathpalette\over@mapsto}

\newcommand{\over@mapsto}[2]{%
   \setbox0=\hbox{$#1#2\mathchar"717F$}%
   \dimen@=\wd0
   \setbox0=\hbox{$#1#2\kern0pt\mathchar"717F$}%
   \advance\dimen@-\wd0
   \vbox{%
      \m@th
      \offinterlineskip
      \ialign{##\cr
              \kern2\dimen@$#1\mapstochar-\mkern-6mu\cleaders\hbox{$#1\mkern-2mu-\mkern-2mu$}\hfill\mkern-7mu{\rightarrow}$\cr
              \hfil$#1#2$\hfil\cr
             }%
    }%
}
\makeatother

\newcommand{\complexity}[1]{Complexity: #1}

\newcommand{\fred}[1]{\textcolor{red}{FL: #1}}
\newcommand{\eric}[1]{\textcolor{blue}{EM: #1}}
\newcommand{\tomasz}[1]{\textcolor{green}{TJ: #1}}

\newcommand{\PM}[1]{P_{#1}}
\newcommand{\SM}[1]{S_{#1}}
\newcommand{\PstarM}[1]{P^{\star}_{#1}}
\newcommand{\SstarM}[1]{S^{\star}_{#1}}
\newcommand{\ILS}[3]{ILS_{#1}(#2)}
\newcommand{\ILSs}[2]{ILS_{#1}(#2)}

\newcommand{\tr}[3]{\tau_{#1,\vv{{#2} {#3}}}}{}
\newcommand{\trs}[4]{P_{#1, \vv{{#2} {#3} \ldots {#4}}}}{}
\newcommand{\trsS}[3]{P_{#1, \vv{{#2}, \ldots, {#3}}}}{}

\newcommand{\trp}[2]{\paths{#1}{q_1}{#2}}{}
\newcommand{\eppaths}[3]{\Pi_{#1,\overline{#2, #3}}}{} 
\newcommand{\ppaths}[2]{\pi_{#1,\overline{#2}}}{} 
\newcommand{\paths}[3]{\rho_{#1,\overline{#2, #3}}}{} 
\newcommand{\pathsp}[3]{p^{+}_{#1,\overline{#2, #3}}}{}
\newcommand{\pathss}[3]{p^{\star}_{#1,\overline{#2, #3}}}{}

\newcommand{\pathsPr}[3]{p_{#1,{\overmapsto{#2, #3}}}}{}
\newcommand{\pathsSu}[3]{p_{#1,\overline{#2, #3}}}{}

\newcommand{\enstrs}[3]{{\mathcal P}_{#1,\vv{{#2},{#3}}}}{}

\newcommand{\delt}[3]{\delta_{#1,\vv{#2, #3}}}{} 
\newcommand{\pdelt}[3]{\Delta_{#1,\vv{#2, #3}}}{} 
\newcommand{\deltp}[3]{\delta^{+}_{#1,\vv{#2, #3}}}{}
\newcommand{\delts}[3]{\delta^{\star}_{#1,\vv{#2, #3}}}{}

\newcommand{\Sp}{S^{+}}{}
\newcommand{\Sm}{S^{-}}{}
\newcommand{\ders}[3]{D_{{#1},\vv{{#2} {#3}}}}{}
\newcommand{\dere}[3]{D^{*}_{{#1},\vv{{#2},{#3}}}}{}

\newcommand{\pref}[3]{{PREF}_{#1,\vv{{#2}~{#3}}}}
\newcommand{\suf}[3]{{SUF}_{#1,\vv{{#2}~{#3}}}}

\newcommand{\nfat}[0]{3NFA}
\newcommand{\knfat}[1]{$#1$\_3NFA}
\newcommand{\nfats}[0]{3NFA$^{\star}$}
\newcommand{\knfats}[1]{$#1$\_3NFA$^{\star}$}

\newcommand{\nffat}[0]{3\_NWFFA}
\newcommand{\knffat}[01]{$#1$\_3NWFFA}

\newcommand{\npfat}[0]{3\_NPFA}
\newcommand{\knpfat}[01]{$#1$\_3NPFA}

\newcommand{\apfs}[1]{a_{#1}^{\star}}

\newcommand{\rpfs}[1]{r_{#1}^{\star}}

\newcommand{\knfa}[0]{$k$\_NFA}
\newcommand{\kpnfa}[0]{$(k+1)$\_NFA}
\newcommand{\knfap}[0]{$k$\_NFA$^+$} 
\newcommand{\kpnfas}[0]{$(k+1)$\_NFA$^\star$} 
\newcommand{\kpnfap}[0]{$(k+1)$\_NFA$^+$} 
\newcommand{\knfas}[0]{$(k+2)$\_NFA$^\star$} 
\newcommand{\knfar}[0]{$(k-1)$\_NFA} 

\newcommand{\kn}[0]{K} %
\newcommand{\kp}[0]{K_{+}} %
\newcommand{\km}[0]{K_{-}} %

\newcommand{\fweight}[2]{\Omega_{(#1,#2)}}
\newcommand{\weight}[3]{\omega_{(#1,#2,#3)}}

\newcommand{\freq}[3]{\phi_{(#1,#2,#3)}}
\newcommand{\afreq}[4]{\phi_{(#1,#2,#3)}(#4)}

\newcommand{\fprob}[2]{\Gamma_{(#1,#2)}}
\newcommand{\prob}[3]{\gamma_{(#1,#2,#3)}}

\newcommand{\fs}[1]{f_{#1}}
\newcommand{\fsp}[1]{f^+_{#1}}
\newcommand{\fss}[1]{\fs{#1}^{*}}

\newcommand{\cmm}{\mathcal{C}_{MM}}
\newcommand{\cma}{\mathcal{C}_{MA}}
\newcommand{\csm}{\mathcal{C}_{SM}}
\newcommand{\csa}{\mathcal{C}_{SA}}


\setcopyright{acmcopyright}

\acmDOI{xx.xxx/xxx_x}

\acmISBN{979-8-4007-0243-3/24/04}

\acmConference[SAC'24]{ACM SAC Conference}{April 8 –April 12, 2024}{Avila, Spain}
\acmYear{2024}
\copyrightyear{2024}

\acmArticle{4}
\acmPrice{15.00}

\title{Classifying Words with 3-sort Automata}
%
\subtitle{Research paper}
%
\renewcommand{\shorttitle}{Classifying Words with 3-sort Automata}
%
\author{Tomasz Jastrz\k{a}b}
\orcid{0000-0002-7854-9058}
\affiliation{%
  \institution{Silesian University of Technology}
  \city{Gliwice} 
  \country{Poland}
}
\email{Tomasz.Jastrzab@polsl.pl}

\author{Fr{\'{e}}d{\'{e}}ric Lardeux}
\orcid{0000-0001-8636-3870}
\affiliation{%
  \institution{LERIA, University of Angers}
  \city{Angers} 
  \country{France}
}
\email{Frederic.Lardeux@univ-angers.fr}

\author{Eric Monfroy}
\orcid{0000-0001-7970-1368}
\affiliation{%
  \institution{LERIA, University of Angers}
  \city{Angers} 
  \country{France}
}
\email{Eric.Monfroy@univ-angers.fr}

\renewcommand{\shortauthors}{T. Jastrz\k{a}b et al.}

\begin{abstract}
Grammatical inference consists in learning a language or a grammar from data. In this paper, we consider a number of models for inferring a non-deterministic finite automaton (NFA) with 3 sorts of states, that must accept some words, and reject some other words from a given sample. We then propose a transformation from this 3-sort NFA into weighted-frequency and probabilistic NFA, and we apply the latter to a classification task. The experimental evaluation of our approach shows that the probabilistic NFAs can be successfully applied for classification tasks on both real-life and superficial benchmark data sets.

\end{abstract}


\maketitle              

\section{Introduction}
Many real-world phenomena may be represented as syntactically structured sequences such as DNA, natural language sentences, electrocardiograms, chain codes, etc. 
Grammatical Inference refers to learning grammars and languages from data, i.e., from such syntactically structured sequences. Machine learning of grammars has various applications in syntactic pattern recognition, adaptive intelligent agents, diagnosis, computational biology, systems modeling, prediction, etc.
We are here interested in learning grammars as finite state automata, with respect to a sample of the language composed of positive sequences that must be elements of the language, and negative sequences that the automaton must reject.

The problem of learning finite automata has been explored from multiple angles: ad-hoc methods such as DeLeTe2~\cite{delete2} which merges states from the prefix tree acceptor (PTA), a family of algorithms for regular language inference presented in~\cite{DBLP:conf/wia/PargaGR06}, metaheuristics such as hill-climbing in~\cite{tomita82}, or modeling the problem as a Constraint Satisfaction Problem (CSP) and solving it with generic tools (such as non-linear programming~\cite{WieczorekBook}, or Boolean formulas~\cite{jastrzab2017,ICCS2023, ICTAI2021,ICTAI2022}).

However, all these works consider Deterministic Finite Automata (DFA), or Non-deterministic Finite Automata (NFA). In both cases, this means that when using the automata on a word, the answer is "Yes, this word is part of the language", or "No, this word is not part of the language". Since samples are finite and usually not very large (a maximum of hundreds of words), and regular languages are infinite, this classification may be too restrictive or too strong. One could be interested in probabilistic answers such as "this word is part of the language with a probability of x\%". However, the question is now "How can I learn a probabilistic automaton from a sample of positive and negative words?".

In this paper, we propose a technique to derive a probabilistic automaton from a sample of positive and negative words. We first learn Non-deterministic Finite Automata with 3 sorts of states: accepting final states which validate positive words, rejecting final states which reject negative words, and whatever states that are not conclusive. We use these 3-sort NFA which seems of reasonable use for our goal (the usefulness of this kind of NFA is presented in~\cite{ColinBook}). To improve the efficiency for generating such automata, we use a similar property to the one that was used for 2-sort NFA in~\cite{ICCS2023}: here, we build a 3-sort automaton with only one accepting final state and one rejecting final state, and some extra constraints to reduce this size $k+2$ automaton into a size $k$ automaton. Then, we want to reflect frequencies based on the sample, such as: how many positive words of the sample terminate in this accepting final state? How many times has a negative word of the sample passed by this transition? But we also want to be a bit more specific. For example, how many negative words of the sample passed by this transition and terminated in a rejecting final state (vs. a whatever state). To this end, we need to weigh differently some cases and patterns. We thus need to define what we call 3-sort Weighted-Frequency NFA, and we present the transformation of a 3-sort NFA into a 3-sort Weighted-Frequency NFA. This last can then be converted into a probabilistic NFA once weights have been instantiated.

The probabilistic NFA can then be used to determine the probability for a word to be a part of the language, or the probability of it not being a part of the language. Note that by modifying the weights, we can obtain more accepting or more rejecting automata. 

We conduct a number of experiments on the Waltz DB database to classify peptides into amyloid ones that are dangerous, or non-amyloid ones that are harmless. We perform similar studies with languages generated by a regular expression. Our results look promising, leaving also some space for weights tuning depending on the aim of the classification, e.g., we can be safer (rejecting dangerous peptides and some harmless ones) or more risky (trying not to reject non-amyloid peptides).

The paper is organized as follows. In Sect.~\ref{sec:nfa} we revise the already developed inference models and propose modifications required to construct 3-sort NFAs. In Sect.~\ref{sec:other-nfa} we show how the 3-sort NFA can be transformed into weighted-frequency NFA and finally into probabilistic NFA. In Sect.~\ref{sec:experiments} we describe conducted experiments and discuss obtained results. Finally, we conclude in Sect.~\ref{sec:conclusion}. 

\section{The NFA inference problem: first models}
\label{sec:nfa}

In this section, we formally present the NFA inference problem based on the propositional logic paradigm and we provide several models. These new models are similar to the ones of~\cite{ICCS2023} but using 3-sort non-deterministic automata.

Without the loss of generality\footnote{If $\lambda \in S$, then it can be recognized or rejected directly, without the need of an automaton.}, we consider in the following, that $\lambda$, the empty word, is not part of the sample. We also consider a unique initial state, $q_1$.

\subsection{Notations}

Let $\Sigma=\{s_1,\ldots,s_n\}$ be an alphabet of $n$ symbols. Moreover, let $\lambda$ denote the empty word, $\kn$ be the set of integers $\{1, \ldots,k\}$, $Pref(w)$ (resp. $Suf(w)$) be the set of prefixes (resp. suffixes) of the word $w$, that we extend to $Pref(W)$ (resp. $Suf(W)$) for a set of words $W$. 



\begin{definition}
A 3-sort non-deterministic finite automaton (\nfat{}) is a 6-tuple ${\mathcal A}=(Q, \Sigma, I, F_{+}, F_{-}, \delta)$ with:
 $Q=\{q_1,\ldots,q_k\}$ -- a finite set of  states,
    $\Sigma$ -- a finite alphabet,
    $I$ -- the set of initial states,
    $F_{+}$ -- the set of accepting final states, $F_{-}$ -- the set of rejecting final states, and $\delta : Q \times \Sigma \rightarrow 2^Q$ -- the transition function. Note that in what follows, we will consider only one initial state, i.e., $I=\{q_1\}$.
\end{definition}

A learning sample $S=\Sp \cup \Sm$ is given by a set $\Sp$ of ``positive'' words from $\Sigma^{*}$ that the inferred 3-sort NFA must accept, and a set $\Sm$ of ``negative'' words that it must reject. 

The language recognized by $\mathcal A$, $L({\mathcal A})_A$, is the set of words for which there exists a sequence of transitions from $q_1$ to a state of $F_{+}$. The language rejected by $\mathcal A$, $L({\mathcal A})_R$, is the set of words for which there exists a sequence of transitions from $q_1$ to a state of $F_{-}$. 

An automaton is non-ambiguous if $L({\mathcal A})_A \cap L({\mathcal A})_R=\emptyset$, i.e., no positive word terminates in a rejecting final state, and no negative word terminates in an accepting final state.

We discard models with 0/1 variables, either from INLP~\cite{WieczorekBook} or CSP~\cite{DBLP:reference/fai/2}: we made some tests with various models with PyCSP~\cite{pycsp3} and obtained some very poor results: the NFA inference problem is intrinsically a Boolean problem, and thus, well suited for SAT solvers.
Hence, we consider the following variables:
\begin{itemize}
 \item $k$, an integer, the size of the \nfat{} to be generated,
 \item a set of $k$ Boolean variables $F_{+}=\{a_1, \ldots, a_k\}$  determining whether state $i$ is a final accepting state or not, 
 \item a set of $k$ Boolean variables $F_{-}=\{r_1, \ldots, r_k\}$  determining if state $i$ is rejecting,
 \item $\Delta=\{\delt{s}{q_i}{q_j}|~s \in \Sigma \textrm{~and~} (i,j) \in {K}^2\}$  , a set of $nk^2$ Boolean variables representing the existence of transitions from state $q_i$ to state $q_j$ with the symbol $s \in \Sigma$, for each $i$, $j$, and $s$.
 
 \item we define $\paths{w}{q_1}{q_{m+1}}$ 
  as the path 
  $q_1,  \ldots, q_{m+1}$ 
  for a word 
  $w=s_1 \ldots  s_m$: 
  $\paths{w}{q_1}{q_{m+1}}=\delt{s_1}{q_1}{q_2} \wedge  \ldots \wedge \delt{s_m}{q_{m}}{q_{m+1}}$.

\end{itemize} 
Although the path is directed from $q_1$ to ${q_{m+1}}$ (it is a sequence of derivations), we will build it either starting from $q_1$, starting from $q_{m+1}$, or starting from both sides \cite{ICTAI2022}. Thus, to avoid confusion, we prefer keeping $\overline{{q_1},{q_{m+1}}}$ without any direction. Paths will be built recursively, and we need at most $\mathcal{O}(\sigma k^{2})$ Boolean variables $\paths{s}{i}{j}$, with $\sigma=\sum_{w \in S}
|w|$.

\subsection{Core of the models}

The core of the models is independent of the way the paths are built. It will thus be common to each model.
The core to define a \nfat{} of size $k$  (noted \knfat{k}) can be defined as follows:   
\begin{itemize}
    \item a final state must be either accepting or rejecting:
       \begin{equation}
         \bigwedge_{i \in K}
         \neg(a_i \wedge r_i)
         \label{final}
        \end{equation}
    
    \item a positive word must terminate in an accepting final state of the \nfat{}, i.e., there must be a path from state $q_1$ to a final state $i$, such that $i \in F_{+}$:
            \begin{equation}
            \bigvee_{i \in K} \paths{w}{q_1}{q_i} \wedge a_i
             \label{posword1}
            \end{equation}

    \item to build a non-ambiguous NFA, a positive word cannot terminate in a rejecting final state:
    \begin{equation}
         \bigwedge_{i \in K} (\neg \paths{w}{q_1}{q_i} \vee \neg r_i)  \label{posword2}
    \end{equation}
    \item similarly, negative words need the same constraints swapping rejecting and accepting:  
    \begin{equation}
        \bigvee_{i \in K} \paths{w}{q_1}{q_i} \wedge r_i  \label{negword1}
    \end{equation}
    \begin{equation}
         \bigwedge_{i \in K} (\neg \paths{w}{q_1}{q_i} \vee \neg a_i)  \label{negword2}
    \end{equation}
\end{itemize}

Of course, the notion of a path can be defined and built in many ways, see \cite{ICTAI2022} for prefix, suffix, and hybrid approaches. 

\subsection{Building paths}

We consider here that a path for a word $w=uv$ is built as the concatenation of a path for the prefix $u$ and one for the suffix $v$. Thus, we can have several joining states ($j$ below) and ending states ($k$ below) for the various paths of a word $w=uv$:
        \begin{eqnarray}
            \bigvee_{(j,k) \in K^2} \paths{u}{q_1}{q_j} \wedge \paths{v}{q_j}{q_k}  \label{prefandsuf}
        \end{eqnarray}

Note that for the empty word $\lambda$ we impose that\footnote{As said before, we consider that $\lambda \not \in S$, but when splitting a word, its prefix or suffix may be $\lambda$.}:
\begin{equation}
 \paths{\lambda}{q_i}{q_j} = True  \texttt{~~~} \forall (i,j) \in K^2 \label{pathLambda},
\end{equation}
to ensure that splitting a word in such a way that the prefix or suffix is the empty word is valid. Note, however, that we do not allow $\lambda$-transitions in the NFA.

Prefixes and suffixes are then built recursively, starting from the beginning of words for prefixes:
\begin{itemize}
    \item For each prefix $u=s$, $s \in \Sigma$: 
\begin{equation}
\bigwedge_{i \in K} \delt{s}{q_1}{q_i} \leftrightarrow \trp{s}{q_i} \label{pref1}
\end{equation}
    \item for each prefix $u=xs$, $s \in \Sigma$ of each word of $S$:
\begin{equation}
\bigwedge_{i \in K} \left(\trp{u}{q_i} \leftrightarrow \left(\bigvee_{j \in K} \trp{x}{q_j} \wedge  \delt{s}{q_j}{q_i}\right)\right) \label{pref2}
\end{equation}
\end{itemize}
and from the ending of words for suffixes:
\begin{itemize}
    \item for $v=s$, and $s \in \Sigma$
    \begin{eqnarray}
    \bigwedge_{(i,j) \in K^2} \delt{s}{q_i}{q_j} \leftrightarrow \paths{s}{q_i}{q_j} \label{suf1}
    \end{eqnarray}
    
    \item for each suffix $v=sx$,  $s \in \Sigma$:
    \begin{eqnarray}
    \bigwedge_{(i,j) \in K^2} \left(\paths{v}{q_i}{q_j} \leftrightarrow 
    \left(\bigvee_{l \in K} \delt{s}{q_i}{q_l}  \wedge  \paths{x}{q_l}{q_j}   \right)\right) \label{suf2}
    \end{eqnarray}
\end{itemize}

\subsection{The models}
\label{models}
We build the models similarly to standard NFA (i.e., without the notion of rejecting states). This means that we have to determine where to split each word $w \in S$ into a prefix $u$ and a suffix $v$. We then consider the set of prefixes $S_u=\{u|w \in S \textrm{~and~} w=uv\}$ and $S_v=\{v|w \in S \textrm{~and~} w=uv\}$ the set of suffixes.
Then, the model is the conjunction of Constraints~(\ref{final})--(\ref{suf2}).

Based on Constraint (\ref{pathLambda}), if we split each word $w$ as $w=w\lambda$, we obtain the prefix model $P$ whose spatial complexity is in ${\mathcal{O}}(\sigma k^{2})$ clauses, and variables, with $\sigma=\sum_{w \in S} |w|$.
Similarly, by splitting words as $w=\lambda w$, we obtain the suffix model $S$ whose spatial complexity is in ${\mathcal{O}}(\sigma k^{3})$ clauses, and variables (the difference is because we do not know where a word terminates).

We then have hybrid models with non-empty suffixes and prefixes. Their complexity is thus in ${\mathcal{O}}(\sigma k^{3})$:
\begin{itemize}
    \item the best suffix model $\SstarM{}$ which consists in determining a minimal set of suffixes covering each word of $S$ and maximizing a cost based on an order over suffixes ($
\Omega(v) = \{ w \in S ~|~ v \textrm{ is a suffix of } w\}
$ and considering two suffixes $v_1$ and $v_2$, 
$
v_1 \succcurlyeq v_2 ~~\Leftrightarrow~~ |v_1|\cdot|\Omega(v_1)| \geq |v_2|\cdot|\Omega(v_2)|
$. Then, prefixes are computed to complete words.

    \item similarly, the best prefix model $\PstarM{}$ is built optimizing prefixes.

    \item we can also try to optimize each word splitting using some metaheuristics, for example, iterated local search (ILS). The model $\ILS{}{Init}{f}$, based on a local search optimization~\cite{Stutzle2018} of word splittings (starting with an initial configuration $Init$, being either a random splitting of words, the splitting found by the $\PstarM{}$ model, or by the $\SstarM{}$ model), tries to minimize the fitness function $f(S_p,S_s)=|Pref(S_p)|+k \cdot |Suf(S_s)|$.

\end{itemize}

\subsection{From $\mathcal{O}(k^3)$ to $\mathcal{O}((k+2)^2)$}
\label{sec:extensions}

Let us consider a sample $S$. If there is a \knfat{k} for $S$, i.e., a \nfat{} of size $k$, to recognize words of $\Sp$ and reject words of $\Sm$, there is also a \knfat{(k+2)} for $S$. We can refine this property by adding some constraints to build what we call \knfas~  extensions. 

Let ${\mathcal A}=(Q, \Sigma, \{q_1\}, F_{+}, F_{-}, \delta)$ be a \knfat{k}. Then, there always exists a \knfat{(k+2)}, ${\mathcal A}'=(Q\cup \{q_{k+1}, q_{k+2}\}, \Sigma, \{q_1\}, \{q_{k+1}\}, \{q_{k+2}\}$, $\delta')$, such that: 

\begin{itemize}
    \item there is only one final accepting state $q_{k+1}$ and one rejecting state $q_{k+2}$, thus we don't need anymore the $a_i$ and $r_i$ variables,

    \item each transition is copied:
 $$
 \begin{array}{l}
 \forall_{i,j \in K^2, s \in \Sigma} \ \delt{s}{q_i}{q_j} \leftrightarrow \delt{s}{q_i}{q_j}'\ ,
 \end{array}
 $$
    \item incoming transitions to accepting final state are duplicated to the new accepting final state $q_{k+1}$:
    $$\begin{array}{l}
 \forall_{i \in K, q_j \in F_{+}} \ \delt{a}{q_i}{q_j} \leftrightarrow \delt{a}{q_i}{q_{k+1}}'\ ,
 \end{array}$$
    \item the same transition duplication is made for rejecting final states to the new rejecting final state $q_{k+2}$,
    \item there is no outgoing transition from states $q_{k+1}$ and $q_{k+2}$ ,
    \item no negative (resp. positive) words terminate in the states from $F_{+}$ (resp. $F_{-}$). This is obvious in $\mathcal{A}$, we have to make it effective in $\mathcal{A'}$.
\end{itemize}

The interest of this \knfat{(k+2)} for $S$ is that the complexity for building suffixes is now in $\mathcal{O}(k^2)$ since both positive and negative words must terminate in a given state (resp. $q_{k+1}$ and $q_{k+2}$).


We now give the constraints of the \knfat{(k+2)}. Let $K_+=\{1,2, \ldots, k+2\}$:
\begin{itemize}
    \item Constraint~(\ref{final}) disappears,
    \item Constraints~(\ref{posword1}) and (\ref{posword2})
    become, for each $w \in \Sp$:
            \begin{equation}
        \paths{w}{q_1}{q_{k+1}} \label{nposword1}
        \end{equation}
        \begin{equation}
          \neg \paths{w}{q_1}{q_{k+2}}   \label{nposword2}
    \end{equation}
    \item the same happens for Constraints~(\ref{negword1})--(\ref{negword2}), replaced by, for $w \in \Sm$:
        \begin{equation}
        \paths{w}{q_1}{q_{k+2}} \label{nnegword1}
        \end{equation}   
        \begin{equation}
        \neg \paths{w}{q_1}{q_{k+1}} \label{nnegword2}
        \end{equation}  

    \item Constraints~(\ref{prefandsuf}) must be split into two, for positive (\ref{nposprefandsuf}) and negative (\ref{nnegprefandsuf}) words:
        \begin{eqnarray}
            \bigvee_{j \in K} \paths{u}{q_1}{q_j} \wedge \paths{v}{q_j}{q_{k+1}}  \label{nposprefandsuf}\\
            \bigvee_{j \in K} \paths{u}{q_1}{q_j} \wedge \paths{v}{q_j}{q_{k+2}}  \label{nnegprefandsuf}
        \end{eqnarray}

    \item Constraints~(\ref{pref1})--(\ref{pref2}) are not modified 

    \item Constraints~(\ref{suf1})--(\ref{suf2}) are respectively modified for positive words into:
    \begin{eqnarray}
    \bigwedge_{i \in K} \delt{s}{q_i}{q_{k+1}} \leftrightarrow \paths{s}{q_i}{q_{k+1}} \label{npossuf1}\\
    \bigwedge_{i \in K} \left(\paths{v}{q_i}{q_{k+1}} \leftrightarrow 
    \left(\bigvee_{j \in K} \delt{s}{q_i}{q_j}  \wedge  \paths{x}{q_j}{q_{k+1}}   \right)\right) \label{npossuf2}
    \end{eqnarray}
    and for negative words into:
    \begin{eqnarray}
    \bigwedge_{i \in K} \delt{s}{q_i}{q_{k+2}} \leftrightarrow \paths{s}{q_i}{q_{k+2}} \label{nnegsuf1}\\
    \bigwedge_{i \in K} \left(\paths{v}{q_i}{q_{k+2}} \leftrightarrow 
    \left(\bigvee_{j \in K} \delt{s}{q_i}{q_j}  \wedge  \paths{x}{q_j}{q_{k+2}}   \right)\right) \label{nnegsuf2}
    \end{eqnarray}

\item There is no outgoing transition from the final states:
\begin{equation}
  \bigwedge_{s \in \Sigma} \bigwedge_{i \in K_+} \neg \delt{s}{q_{k+1}}{q_i} ~\wedge~ \neg \delt{s}{q_{k+2}}{q_i}\label{nooutgoing}
\end{equation}

\item Each incoming transition of the  accepting (resp. rejecting) final state $q_{k+1}$ (resp. $q_{k+2}$) must also terminate in another state (duplication):
\begin{eqnarray}
  \bigwedge_{s \in \Sigma}
    \Bigg(
        \bigwedge_{~i \in K} 
            \bigg(
                \delt{s}{q_i}{q_{k+1}} \rightarrow \bigvee_{j \in K} \delt{s}{q_i}{q_j}
            \bigg)
    \Bigg)
    \label{dup1}\\
  \bigwedge_{s \in \Sigma}
    \Bigg(
        \bigwedge_{~i \in K} 
            \bigg(
                \delt{s}{q_i}{q_{k+2}} \rightarrow \bigvee_{j \in K} \delt{s}{q_i}{q_j}
            \bigg)
    \Bigg)        \label{dup2}
\end{eqnarray}
\end{itemize}

In order to be able to reduce the \knfats{(k+2)} into a \knfat{k}, we must take care about the possibly rejecting and accepting final states of the \knfat{k}.
To this end, we need a new set of Boolean variables representing possibly accepting (resp. rejecting) final states for the corresponding \knfa{}: $\{\apfs{1}, \cdots, \apfs{k}\}$ (resp. $\{\rpfs{1}, \cdots, \rpfs{k}\}$). The \knfats{(k+2)} may be reduced to a \knfat{k} by just removing states $q_{k+1}$, $q_{k+2}$, and their incoming transitions, and by fixing the final states among the possible final states, i.e., determining the $\apfs{i}$ and the 
$\rpfs{i}$  which are final states of the \knfat{k}, either accepting or rejecting. To determine these possible final states, we have to ensure:
\begin{itemize}
\item A negative (resp. positive) word cannot terminate in an accepting (resp. rejecting) possible final state:
\begin{eqnarray}
  \bigwedge_{i \in K} \bigg(\apfs{i} \rightarrow \bigwedge_{w \in \Sm} \neg \paths{w}{q_1}{q_i} \bigg) \label{fsnonegword}\\
    \bigwedge_{i \in K} \bigg(\rpfs{i} \rightarrow \bigwedge_{w \in \Sp} \neg \paths{w}{q_1}{q_i} \bigg) \label{fsnoposword}
\end{eqnarray}
Note that with Constraints~(\ref{fsnonegword})--(\ref{fsnoposword}), Constraints (\ref{posword2})--(\ref{negword2}) are no longer needed.

\item Each accepting (resp. rejecting) possible final state validates at least one positive (resp. negative) word of $S$:
\begin{eqnarray}
  \bigwedge_{i \in K} \left(\apfs{i} \rightarrow \left(\bigvee_{vs \in \Sp} \bigvee_{~j \in K} \left(\paths{v}{q_1}{q_j} \wedge  \delt{s}{q_j}{q_i}  \wedge  \delt{s}{q_j}{q_{k+1}}\right)\right)\right)\label{fsposword}\\
  \bigwedge_{i \in K} \left(\rpfs{i} \rightarrow \left(\bigvee_{vs \in \Sm} \bigvee_{~j \in K} \left(\paths{v}{q_1}{q_j} \wedge  \delt{s}{q_j}{q_i}  \wedge  \delt{s}{q_j}{q_{k+2}}\right)\right)\right)\label{fsnedword}  
\end{eqnarray}

\item Each positive (resp. negative) word terminates in at least one accepting (resp. rejecting) possible final state:
\begin{eqnarray}
  \bigwedge_{w \in \Sp}  \bigvee_{~i \in K} (\paths{w}{q_1}{q_i} \wedge \apfs{i}) \label{posword}\\
  \bigwedge_{w \in \Sm}  \bigvee_{~i \in K} (\paths{w}{q_1}{q_i} \wedge \rpfs{i}) \label{negword}  
\end{eqnarray}

\item a state cannot be both accepting and rejecting possible final state:
\begin{eqnarray}
  \bigwedge_{i \in K}  
  \neg(\apfs{i} \wedge \rpfs{i}) \label{eitherpossible}
\end{eqnarray}

\end{itemize}



\section{From \nfat{} to weighted-frequency NFA and probabilistic NFA}\label{sec:other-nfa}

Using a sample of positive and negative words, we are able to generate a \knfat{k}. However, we cannot directly obtain probabilistic automata to decide the probability for a word to be part or not of the language represented by the sample. However, we can use the sample and the generated \knfat{k} to build a weighted-frequency automaton: weighted-frequencies will be determined with respect to the sample words. We can then create, from this last automaton, a probabilistic automaton to classify words.

\subsection{Weighted-frequency automata}
We now define what we call a weighted-frequency automaton. In a frequency automata, the integer $n$ attached to a transition $\delta(q,a,q')$ means that this transition was used $n$ times (see~\cite{ColinBook}). Here, we want to count differently positive (resp. negative) words terminating in an accepting (resp. rejecting) final state from positive (resp. negative) words terminating in whatever state. We thus need to weigh these cases with different real numbers. Thus, we obtain automata that reflect weighted frequencies and are still based on 3-sort automata.

\begin{definition}[\nffat{}]
A 3-sort non-deterministic weighted-\-frequency finite automaton (\nffat{}) is a 10-tuple ${\mathcal A}=(Q, \Sigma, I, F_{+},$ $F_{-}, \delta, \fweight{f}{+}, \fweight{f}{-}, \fweight{\delta}{+}, \fweight{\delta}{-})$  with:
\begin{itemize}
    \item $Q=\{q_1,\ldots,q_k\}$ -- a finite set of  states,

    \item $\Sigma$ -- a finite alphabet,

    \item $I=\{q_1\}$ -- the set of initial states,

    \item $F_{+}$ -- the set of accepting final states, 
    
    \item $F_{-}$ -- the set of rejecting final states, 

    \item $\delta : Q \times \Sigma \rightarrow 2^Q$ -- the transition function,



    \item $\fweight{f}{+}$ -- a function $Q \rightarrow \mathbb{N}$, i.e., $\fweight{f}{+}(q)=$
    $$
    \left\{
    \begin{array}{lcl}
    \weight{f}{+}{+}\cdot\afreq{f}{+}{+}{q} & \textrm{if}  & q \in F_{+} \\ 
    \weight{f}{+}{?}\cdot\afreq{f}{+}{?}{q} & \textrm{if} & q \in Q \setminus (F_{+} \cup F_{-}) \\
    0  & &\textrm{otherwise}
    \end{array}
    \right.
    $$
    where:
    \begin{itemize}
        \item $\weight{f}{+}{+}$ and $\weight{f}{+}{?}$ two weights associated respectively to positive words terminating in accepting states and to positive words terminating in whatever states,
        \item two counting functions $\afreq{f}{+}{+}{q}$ and $\afreq{f}{+}{?}{q}$ for counting the number of times (i.e., the numbers of physical paths for all positive words) a positive word terminates in accepting state $q$ and respectively in whatever state $q$. These two functions are detailed later.
    \end{itemize}

    \item $\fweight{f}{-}$ -- a function $Q \rightarrow \mathbb{N}$, $\fweight{f}{-}(q)$
    is defined as above but for negative words and rejecting states (i.e., "+" is replaced by "-", and $F_{+}$ by $F_{-}$).

    \item $\fweight{\delta}{+}$  -- a function $Q \times \Sigma \times Q \rightarrow \mathbb{N}$, i.e., $\fweight{\delta}{+}(q,s,q')=$
    $$\weight{\delta}{+}{+}\cdot\afreq{\delta}{+}{+}{q,s,q'}
    +
    \weight{\delta}{+}{?}\cdot\afreq{\delta}{+}{?}{q,s,q'}$$ 
    where:
    \begin{itemize}
        \item $\weight{\delta}{+}{+}$ and $\weight{\delta}{+}{?}$ are two weights associated respectively with positive words terminating in accepting states and positive words terminating in whatever states,
        \item two counting functions $\afreq{\delta}{+}{+}{q,s,q'}$ and $\afreq{\delta}{+}{?}{q,s,q'}$ for counting the number of times a positive word uses given transition within a physical path that terminates in an accepting state and respectively in whatever state. These two functions are detailed later.
    \end{itemize}

    \item $\fweight{\delta}{-}$  -- a function $Q \times \Sigma \times Q \rightarrow \mathbb{N}$, i.e., $\fweight{\delta}{-}(q,s,q')=\weight{\delta}{-}{-}\cdot\afreq{\delta}{-}{-}{q,s,q'}
    +
    \weight{\delta}{-}{?}\cdot\afreq{\delta}{-}{?}{q,s,q'}$   
    defined as above but for negative words.
\end{itemize}

Remember that these automata are non-deterministic, and thus, there can be several paths for a word, terminating in different states. Thus, for a given word, we are interested in all terminating paths independently from the sort of the terminating state.

\subsection{From 3-sort NFA to weighted-frequency automata}
We need a ``physical'' view of transitions and paths of the \knfat{k} we have built. Consider the transition function $\delta: Q \times \Sigma \rightarrow 2^Q$ from a \knfat{k} $\mathcal{A}$. 
We rename by $\pdelt{s}{i}{j}$ the value $\delta(q_i,s,q_j)$. Note that if $\delt{s}{q_i}{q_j}$ is true, $\pdelt{s}{i}{j}$ exists, otherwise $\pdelt{s}{i}{j}$ does not exist. 

We also define $\ppaths{s_1 \ldots s_n}{i_1, \ldots, i_{n+1}}$ as the sequence of physical transitions $\pdelt{s_1}{i_1}{i_2}, \ldots, \pdelt{s_n}{i_{n}}{i_{n+1}}$.
For a given word $w=s_1 \ldots s_n$, 
$$\eppaths{w}{i_1}{i_{n+1}}=\{\ppaths{s_1 \ldots  s_n}{i_1, \ldots, i_{n+1}}~|~i_1, \ldots, i_{n+1} \in Q^{n+1}\}$$
is the set of all sequences for $w$ in $\mathcal{A}$. 

Consider a sequence $\ppaths{s_1 \ldots s_n}{i_1, \ldots, i_{n+1}}=\pdelt{s_1}{i_1}{i_2}, \ldots, \pdelt{s_n}{i_{n}}{i_{n+1}}$. Then,  $occ(\ppaths{s_1 \ldots s_n}{i_1, \ldots, i_{n+1}})(\pdelt{s}{i}{j})$ is the number of occurrences of $\pdelt{s}{i}{j}$  in the sequence $\ppaths{s_1. \ldots. s_n}{i_1, \ldots, i_{n+1}}$
defined recursively as follows:
\begin{itemize}
    \item $occ(\Lambda)(\pdelt{s}{i}{j})=0$
    \item $occ(\pdelt{s_1}{k_1}{l_1} , \pdelt{s_2}{k_2}{l_2}, \ldots, \pdelt{s_3}{k_3}{l_3})(\pdelt{s}{i}{j})=$
    $$
    \left\{
    \begin{array}{l}
    1+occ(\pdelt{s_2}{k_2}{l_2}, \ldots, \pdelt{s_3}{k_3}{l_3})(\pdelt{s}{i}{j}) 
        \textrm{ if } (s,i,j)=(s_1,k_1,l_1),\\
    occ(\pdelt{s_2}{k_2}{l_2}, \ldots, \pdelt{s_3}{k_3}{l_3})(\pdelt{s}{i}{j})
        \textrm{ otherwise}
    \end{array}
    \right.
    $$.
\end{itemize}

We now propose a possible implementation of the counting functions for a weighted-frequency automaton:
\begin{itemize}
    \item $\freq{f}{+}{+}$: if $q \in F_{+}$, $\afreq{f}{+}{+}{q}=|\bigcup_{w \in \Sp} \eppaths{w}{1}{q}|$, 0 otherwise

    \item $\freq{f}{+}{?}$: if $q \in Q \setminus (F_{+} \cup F_{-})$, $\afreq{f}{+}{?}{q}=|\bigcup_{w \in \Sp} \eppaths{w}{1}{q}|$, 0 otherwise

    \item $\freq{f}{-}{-}$ and $\freq{f}{-}{?}$ can be defined similarly 

    \item $\freq{\delta}{+}{+}$: $\freq{\delta}{+}{+}(q,s,q')=$
    $$
    \sum_{w \in \Sp} \bigg(
    \sum_{q_A \in F_{+}} \bigg(
    \sum_{p \in \eppaths{w}{q_1}{q_A}}
    occ(p)(\pdelt{s}{q}{q'})
    \bigg) \bigg)
    $$

    \item $\freq{\delta}{+}{?}$, $\freq{\delta}{-}{-}$, and $\freq{\delta}{-}{?}$ are defined similarly.
\end{itemize}
We can imagine other counting functions, for example not considering all possible paths for a word, but only one path, or only a given number of paths.

The different weights enable us to consider only positive words for example ($\weight{F}{-}{*}=0$ for $F \in \{f, \delta\}$ and $*=\{-,?\}$), or considering for example only positive words terminating in a positive state ($\weight{f}{+}{?}=0$, and $\weight{\delta}{+}{?}=0$).
\end{definition}

%

\subsection{Probabilistic automata}

We can now define the probabilistic automata we are interested in: 3-sort automata with probabilities for transitions and probabilities for states to be final accepting and rejecting.

\begin{definition}[\npfat{}]
A 3-sort non-deterministic probabilistic finite automaton is 
 is an 8-tuple ${\mathcal A}=(Q, \Sigma, I, \delta, \fprob{f}{+}, \fprob{f}{-},$ $\fprob{\delta}{+}, \fprob{\delta}{-})$ with:
\begin{itemize}
    \item $Q=\{q_1,\ldots,q_k\}$ -- a finite set of  states,

    \item $\Sigma$ -- a finite alphabet,

    \item $I=\{q_1\}$ -- the set of initial states,

    \item $\delta : Q \times \Sigma \rightarrow 2^Q$ -- the transition function,



    \item $\fprob{f}{+}$ -- a function $Q \rightarrow [0,1]$, i.e., $\fprob{f}{+}(q)$ is the probability of state $q$ to be accepting final,

    \item $\fprob{f}{-}$ -- a function $Q \rightarrow  [0,1]$, i.e., $\fprob{f}{-}(q)$ is the probability of state $q$ to be rejecting final,

    \item $\fprob{\delta}{+}$ -- a function $Q \times \Sigma \times Q \rightarrow [0,1]$, i.e., $\fprob{\delta}{+}(q,s,q')$ is the probability for a positive word to pass by the transition $\delta(q,s,q')$,

    \item $\fprob{\delta}{-}$ -- similar to $\fprob{\delta}{+}$ for negative words.

\end{itemize}
A 3-sort non-deterministic probabilistic automaton  ${\mathcal A}=(Q, \Sigma, I, \delta,$ $\fprob{f}{+}, \fprob{f}{-}, \fprob{\delta}{+}, \fprob{\delta}{-})$ must respect the following constraint:
$$
\forall q \in Q,
\left\{
\begin{array}{lcl}
\sum_{q' \in Q, s \in \Sigma} 
    \big( \fprob{\delta}{+}(q,s,q') + \fprob{f}{+}(q) \big) = 1\\
\sum_{q' \in Q, s \in \Sigma} 
    \big( \fprob{\delta}{-}(q,s,q') + \fprob{f}{-}(q) \big) = 1
\end{array}
\right.
$$
\end{definition}

Remember that we consider only one initial state which is $q_1$. In case one wants to consider several initial states, some probabilities of being initial positive and initial negative can be added.

\subsection{From weighted-frequency to probabilistic automata}
We now present the transformation of a \nffat{} into a \npfat{}: weighted-frequencies are converted into probabilities. 



Consider a 3-sort non-deterministic weighted-frequency finite automaton ${\mathcal A}=(Q, \Sigma, I, F_{+},$ $F_{-}, \delta, \fweight{f}{+}, \fweight{f}{-}, \fweight{\delta}{+}, \fweight{\delta}{-})$. Then, from ${\mathcal A}$, we can derive a 3-sort non-deterministic probabilistic finite automaton ${\mathcal A'}=(Q', \Sigma', I', \delta', \fprob{f}{+}, \fprob{f}{-}, \fprob{\delta}{+}, \fprob{\delta}{-})$ such that:
\begin{itemize}
    \item the states, alphabet, transitions, and initial state remain unchanged:    
    $Q=Q'$, $\Sigma=\Sigma'$, $I=I'$, and $\delta=\delta'$

    
    
    \item the probability for $q$ to be an accepting final state is the weighted frequency of words of $\Sp$ terminating in $q$, divided by the sum of the weighted frequencies of the positive words from the sample outgoing from $q$ plus the weighted-frequency of positive words ending in $q$:    
    $$
    \begin{array}{l}
        \forall q \in Q,~
                \fprob{f}{+}(q)=  \\
        \hspace*{10mm}
        \fweight{f}{+}(q) 
                / 
                \bigg(
                \fweight{f}{+}(q) + 
                \sum_{s \in \Sigma, q' \in Q} \fweight{\delta}{+}(q,s,q')
                \bigg) 
    \end{array}
    $$
    
    
    \item the probabilities $\fprob{f}{-}$ are computed similarly for negative words replacing "+" by "-".
        
    \item the probability for a positive word to follow transition $\delta(q,s,q')$ is computed similarly as the probability of ending in $q$:
    $$\begin{array}{l}
        \forall q, q' \in Q, \forall s \in \Sigma, \fprob{\delta}{+}(q,s,q')= ~\\
        \hspace*{8mm}
        \fweight{\delta}{+}(q,s,q') 
                / 
                \bigg(
                \fweight{f}{+}(q)
                +
                \sum_{s' \in \Sigma, q'' \in Q} \fweight{\delta}{+}(q,s',q'')
                \bigg)
    \end{array}
    $$ 
   
    \item the computation is similar for negative words replacing "+" by "-".   
\end{itemize}
These probabilities respect the properties of 3-sort non-deterministic frequency finite automata.


\subsection{Classifying words}

Given the two sets of independent weights (for states and transitions) and the non-deterministic nature of the NFA, implying possibly multiple paths for a word, we consider four classifiers:
\begin{itemize}
    \item $\cmm$ -- computes the positive and negative scores for a word by \textit{multiplying} the probabilities of the transitions and the probability of the last state on each path, selecting as the final score the \textit{maximum} of all paths.
    \item $\cma$ -- computes the positive and negative scores for a word by \textit{multiplying} the probabilities of the transitions and the probability of the last state on each path, selecting as the final score the \textit{average} of all paths.
    \item $\csm$ -- computes the positive and negative scores for a word by \textit{summing up} the probabilities of the transitions and the probability of the last state on each path, selecting as the final score the \textit{maximum} of all paths. Summed up probabilities for each path for word $w$ are divided by $|w| + 1$, to scale them to the range $[0, 1]$.
    \item $\csa$ -- computes the positive and negative scores for a word by \textit{summing up} the probabilities of the transitions and the probability of the last state on each path, selecting as the final score the \textit{average} of all paths. Summed up probabilities for each path for word $w$ are divided by $|w| + 1$, to scale them to the range $[0, 1]$.
\end{itemize}
The final classifier decision, i.e., acceptance or rejection of a word, is based on the comparison of positive and negative scores---the greater score wins.

To illustrate the operation of the classifiers let us consider an example. Assume that we have a word $w = abb$ for which there are two paths in some NFA. For simplicity, assume that all weights $\weight{F}{\bullet}{*} = 1$, with $F = \{f, \delta\}$, $\bullet = \{+, -\}$, and $* = \{+, -, ?\}$. Let us also assume that the transition and last state probabilities for the first path are: $(0.2, 0.5, 0.35, 0.6)$ for the acceptance scenario, and $(0.6, 0.5, 0.65, 0.4)$ for the rejection scenario. For the second path assume probabilities $(0.2, 0.15, 0.55, 0.9)$ and $(0.6, 0.5, 0.5, 0.75)$. Then the scores for the respective classifiers are as follows:
\begin{itemize}
    \item for $\cmm$ the positive score is 0.02, and the negative is 0.11,
    \item for $\cma$ the positive score is 0.02, and the negative is 0.10,
    \item for $\csm$ the positive score is 0.45, and the negative is 0.59,
    \item for $\csa$ the positive score is 0.43, and the negative is 0.56.
\end{itemize}
It is clear that each classifier indicates that the word should be rejected as the negative scores are always greater than the positive ones. Also, we can note that the $\csm$ and $\csa$ produce larger margin between the scores then the $\cmm$ and $\cma$ (0.13--0.14 vs. 0.08--0.09).


\section{Experimentation}
\label{sec:experiments}

\subsection{Experiment I}

To evaluate the proposed probabilistic automata and classifiers, we have created a benchmark set based on the peptides stored in WaltzDB database \cite{waltzdb,waltzdb1}. The benchmark set was composed of several samples containing amyloid (positive) and non-amyloid (negative) peptides, each having a length of 6 characters. The samples were created based on peptide subsets available on the WaltzDB website (\url{http://waltzdb.switchlab.org/sequences}).

Based on each sample, the training and test samples were created, with the training sample consisting of 10\%, 30\%, and 50\% of the first peptide sequences in the given subset. The training sample was used to infer the probabilistic NFA, which acted then as a classifier for the test sample, comprising the whole subset without the elements included in the training sample. Since some of the subsets contained very few positive/negative sequences, for the final evaluation we selected only five of them, i.e., Amylhex (AH), Apoai mutant set (AMS), Literature (Lit), Newscores (NS), and Tau mutant set (TMS). Table~\ref{tab:dataset} summarizes the characteristics of the data set. Note that all samples are quite imbalanced and they differ both in the total number of words and the size of the alphabet.

\begin{table}[h]
    \centering
    \caption{Characteristics of the benchmark set}
    \label{tab:dataset}
    \resizebox{\columnwidth}{!}{
    \begin{tabular}{lr|rr|rr|rr|rr}
    \toprule
         & & \multicolumn{2}{c|}{Train 10\%}& \multicolumn{2}{c|}{Train 30\%} & \multicolumn{2}{c|}{Train 50\%} & \multicolumn{2}{c}{Whole subset}\\
        Subset & \multicolumn{1}{c|}{$|\Sigma|$} & \multicolumn{1}{c}{$|\Sp|$} & \multicolumn{1}{c|}{$|\Sm|$} & \multicolumn{1}{c}{$|\Sp|$} & \multicolumn{1}{c|}{$|\Sm|$} & \multicolumn{1}{c}{$|\Sp|$} & \multicolumn{1}{c|}{$|\Sm|$} & \multicolumn{1}{c}{$|\Sp|$} & \multicolumn{1}{c}{$|\Sm|$}\\
        \midrule
        AH & 19 & 7 & 12 & 23 & 36 & 39 & 60 & 79 & 121 \\
        AMS & 20 & 7 & 3 & 23 & 10 & 39 & 18 & 79 & 36 \\
        Lit & 20 & 20 & 6 & 61 & 19 & 102 & 33 & 204 & 66 \\
        NS & 18 & 3 & 1 & 9 & 4 & 16 & 7 & 32 & 15 \\
        TMS & 19 & 9 & 2 & 27 & 6 & 46 & 11 & 92 & 22 \\
        \bottomrule
    \end{tabular}
    }
\end{table}

The inference models were implemented in Python using PySAT library and the Glucose SAT solver with default options. The experiments were carried out on a computing cluster with Intel-E5-2695 CPUs, and a fixed limit of 10 GB of memory. Running times were limited to 15 minutes, including model generation and solving time. The classification was conducted using a Java application running on a single machine with Intel Core i7-7560U 2.40GHz processor and 8 GB of RAM. 






Since weights tuning lies outside of the scope of the current paper, we decided to conduct the experiments by setting respective weights to 0s or 1s only, analyzing all possible combinations of 0s and 1s for the eight weights defined before. Thus, for each training sample we analyzed 256 different weight assignments, which along with 115 inferred NFAs\footnote{In five cases for the NS subset, we failed to infer an NFA for the Train 50\% training set. The models that failed were the $\PstarM{k+2}$ and all suffix-based models.} and 4 classifiers gave us a total of 117\hspace{3pt}760 classifications. The whole process took around 186 minutes, with the \knfat{(k+2)} models taking on average 2.9 times longer to perform the classification than the \knfat{k} ones. This difference may be attributed to the larger size of the former NFAs, which makes the path building process more time-consuming.

The classification results were evaluated based on accuracy and F1-score given by Eqs. (\ref{eqn:accuracy}) and (\ref{eqn:f1-score}):
\begin{equation}\label{eqn:accuracy}
    Acc = \frac{TP + TN}{TP + TN + FP + FN}
\end{equation}
\begin{equation}\label{eqn:f1-score}
    F1 = \frac{2\cdot TP}{2\cdot TP + FP + FN}
\end{equation}
where $TP$ denotes true positives (amyloid peptides classified as amyloid), $TN$ denotes true negatives (non-amyloid peptides classified as such), $FP$ denotes false positives (non-amyloid peptides classified as amyloid ones), and $FN$ denotes false negatives (amyloid peptide classified as non-amyloid).


Table~\ref{tab:res-models-subsets} shows the best accuracy values and their corresponding F1-scores obtained for the test sets over all analyzed weight combinations and all classifiers. The metrics were obtained by NFAs inferred using 8 different models. Boldfaced values denote the best column-wise values. The entries with an asterisk denote the cases in which the best F1-score did not correspond to the best accuracy.


\begin{table}[htb]
    \centering
    \caption{Best accuracy and corresponding F1-score metrics obtained by the NFAs for the analyzed benchmark sets}
    \label{tab:res-models-subsets}
    \resizebox{\columnwidth}{!}{
    \begin{tabular}{l|rrrrr|rrrrr}
    \toprule
    & \multicolumn{5}{c|}{Accuracy}& \multicolumn{5}{c}{F1-score}\\
    Model & \multicolumn{1}{c}{AH} & \multicolumn{1}{c}{AMS} & \multicolumn{1}{c}{Lit} & \multicolumn{1}{c}{NS} & \multicolumn{1}{c|}{TMS} & \multicolumn{1}{c}{AH} & \multicolumn{1}{c}{AMS} & \multicolumn{1}{c}{Lit} & \multicolumn{1}{c}{NS} & \multicolumn{1}{c}{TMS} \\
    \midrule
    $\PM{k}$ & 0.63 & 0.59 & 0.76 & 0.63 & 0.72 & 0.63* & 0.66* & \textbf{0.86} & 0.73 & \textbf{0.82}\\
    $\PM{(k+2)}$ & \textbf{0.69} & \textbf{0.68} & 0.76 & 0.67 & \textbf{0.73} & 0.64* & \textbf{0.79} & \textbf{0.86} & 0.76 & \textbf{0.82}\\
    $\PstarM{k}$ & 0.68 & 0.62 & 0.76 & \textbf{0.71} & 0.72 & \textbf{0.66} & 0.72 & \textbf{0.86} & \textbf{0.77} & \textbf{0.82}\\
    $\PstarM{(k+2)}$ & 0.64 & 0.67 & 0.73 & 0.50 & \textbf{0.73} & 0.62* & 0.77 & 0.82 & 0.51 & \textbf{0.82}\\
    $\SM{k}$ & 0.61 & 0.55 & 0.76 & 0.62 & 0.72 & 0.48* & 0.56* & \textbf{0.86} & 0.70 & \textbf{0.82}\\
    $\SM{(k+2)}$ & 0.65 & 0.64 & \textbf{0.77} & 0.50 & \textbf{0.73} & 0.60* & 0.77 & \textbf{0.86} & 0.48 & \textbf{0.82}\\
    $\SstarM{k}$ & 0.61 & 0.62 & 0.66 & 0.50 & 0.72 & 0.58* & 0.76 & 0.79 & 0.51 & \textbf{0.82}\\
    $\SstarM{(k+2)}$ & 0.63 & 0.66 & 0.74 & 0.47 & \textbf{0.73} & 0.42* & 0.77 & 0.85 & 0.44 & \textbf{0.82}\\
    \bottomrule
    \end{tabular}
    }
\end{table}

Based on the accuracy values we can state that the prefix-based models perform best among all eight models, regardless of the benchmark set. It is also clear that NS data set turned out to be the hardest one, since some of the models achieved accuracy smaller than 0.5, which is the probability of success with a random decision in binary classification. The analysis of F1-score, being the harmonic mean between precision and recall, confirms the strong position of prefix-based models, with a small advantage given to the $\PstarM{k}$ model. Overall, the achieved metrics are not very satisfactory, which may be attributed to, e.g.:
\begin{itemize}
    \item the way the training samples were constructed -- it is not infrequent that the training sample does not cover the whole alphabet, which results in the rejection of words from the test set using the symbols outside of the training sample's alphabet,
    \item the lack of some language behind the subsets of peptides -- there is no guarantee that the peptides from a certain subset share some common features reflected in the sequences,
    \item limited parameter tuning -- so far we analyzed only the extreme values for the weights, more advanced parameter tuning may be required to achieve better results.
\end{itemize}

Interestingly, the best results were typically achieved with 30\% training sets. The NS data set, whenever the NFA for it could be inferred, required the 50\% training set to achieve its peak performance. There were also rare cases in which the smallest training data set was sufficient (e.g., for the $\PM{k}$ model with AMS data set).

Table~\ref{tab:res-models-classifiers} shows the best accuracy values and corresponding F1-scores for different classifiers over all analyzed weight combinations and all subsets. The metrics pertain to NFAs inferred using eight different models. The meaning of boldfaced entries and entries with an asterisk is the same as for Table~\ref{tab:res-models-subsets}.

\begin{table}[htb]
    \centering
    \caption{Best accuracy and corresponding F1-score metrics obtained by the NFAs for the analyzed classifiers}
    \label{tab:res-models-classifiers}
    \resizebox{\columnwidth}{!}{
    \begin{tabular}{l|rrrr|rrrr}
    \toprule
    & \multicolumn{4}{c|}{Accuracy} & \multicolumn{4}{c}{F1-score}\\
    Model & \multicolumn{1}{c}{$\cmm$} & \multicolumn{1}{c}{$\cma$} & \multicolumn{1}{c}{$\csm$} & \multicolumn{1}{c|}{$\csa$} & \multicolumn{1}{c}{$\cmm$} & \multicolumn{1}{c}{$\cma$} & \multicolumn{1}{c}{$\csm$} & \multicolumn{1}{c}{$\csa$} \\
    \midrule
    $\PM{k}$ & \textbf{0.70} & \textbf{0.70} & 0.76 & 0.76 & \textbf{0.80} & \textbf{0.81} & \textbf{0.86} & \textbf{0.86}\\
    $\PM{(k+2)}$ & \textbf{0.70} & 0.69 & 0.76 & 0.76 & \textbf{0.80} & 0.79 & \textbf{0.86} & \textbf{0.86}\\
    $\PstarM{k}$ & 0.56 & 0.56 & 0.76 & 0.76 & 0.45* & 0.45* & \textbf{0.86} & \textbf{0.86}\\
    $\PstarM{(k+2)}$ & 0.59 & 0.59 & 0.73 & 0.73 & 0.70 & 0.70 & 0.82* & 0.82*\\
    $\SM{k}$ & 0.53 & 0.53 & 0.76 & 0.76 & 0.61* & 0.69 & \textbf{0.86} & \textbf{0.86}\\
    $\SM{(k+2)}$ & 0.64 & 0.64 & \textbf{0.77} & \textbf{0.77} & 0.73 & 0.73 & \textbf{0.86} & \textbf{0.86}\\
    $\SstarM{k}$ & 0.53 & 0.53 & 0.72 & 0.72 & 0.69 & 0.69 & 0.82 & 0.82\\
    $\SstarM{(k+2)}$ & 0.68 & 0.68 & 0.74 & 0.74 & 0.79 & 0.79 & 0.85 & 0.85\\
    \bottomrule
    \end{tabular}
    }
\end{table}

The analysis shows that this time $\PM{k}$ is clearly the best model. In terms of classifiers, we do not observe many differences between the pairs of classifiers based on multiplication ($\cmm$, $\cma$) and summation ($\csm$, $\csa$). However, the differences between multiplication- and summation-based classifiers for the given model are statistically significant (based on ANOVA test with $\alpha = 0.05$ and post hoc Tukey HSD) in terms of both accuracy and F1-score.

\begin{figure}[htb]
    \centering
    \includegraphics[width=\columnwidth]{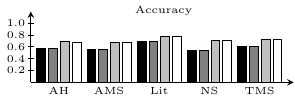}
    \\
    \includegraphics[width=\columnwidth]{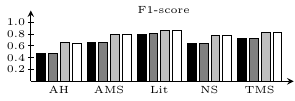}
    \caption{Best accuracy and F1-score metrics obtained by all NFAs for the analyzed benchmark sets and classifiers $\cmm$ (black), $\cma$ (gray), $\csm$ (light gray), and $\csa$.}
    \label{fig:res-subsets-classifiers}
\end{figure}

Figure~\ref{fig:res-subsets-classifiers} shows the best accuracy values and corresponding F1-scores obtained by all classifiers over all analyzed weight combinations and NFAs inferred by all models. We can confirm that the differences between the classifiers are statistically significant at $\alpha = 0.05$, with the $\csa$ and $\csm$ classifiers consistently achieving better results than the other two. We can also note that Lit data set was the most favorable in terms of satisfactory metric values. 

\subsection{Experiment II}

To evaluate the proposed solutions even further, we created the second benchmark composed of two data sets. The data sets were built based on regular expressions (regexp)\footnote{The actual regular expressions were as follows: (0|11)(001|000|10)*0 and [0-9][0-4][5-9](024|135|(98|87))*(0|6).}---we defined two languages described by different regular expressions from which we sampled the words of 1 to 15 characters. These words represented the sets $\Sp$. The sets $\Sm$ contained words constructed by randomly shuffling the positive examples and ensuring they do not match the regexp. Similarly to the first experiment, we created the training data sets used for NFA inference and the test sets for evaluation. The sizes of complete samples were equal to 200 words split equally between $\Sp$ and $\Sm$. The experimental setup, i.e., computing machines, metrics, and weight settings were kept as in Experiment I.

In Table~\ref{tab:res-models-subsets2} we show the results obtained for the various models across all classifiers for the two regexp-based data sets. Comparing the results to the ones presented in Tab.~\ref{tab:res-models-subsets}, we note a significant improvement in the achieved metrics. We also observe that for RegExp1 data set, except for the $\SM{k}$ model, all models achieve perfect scores. Finally, we note that for RegExp2 all $(k+2)$-based models, except $PM{(k+2)}$, improve over their $k$-based counterparts, while for RegExp1 it only applies to $\SstarM{}$ model, since the others achieved perfect scores. Detailed analysis have shown that in most cases, the best accuracy and F1-score were obtained with the 50\% training set, but for the $\PM{(k+2)}$, $\SM{k}$, and $\SstarM{(k+2)}$ models, 10\% was enough.

\begin{table}[htb]
    \centering
    \caption{Best accuracy and corresponding F1-score metrics obtained by the NFAs for the analyzed benchmark sets.}
    \label{tab:res-models-subsets2}
    \resizebox{\columnwidth}{!}{
    \begin{tabular}{ll|rrrrrrrr}
    \toprule
    & Data set& \multicolumn{1}{c}{$\PM{k}$}& \multicolumn{1}{c}{$\PM{(k+2)}$}& \multicolumn{1}{c}{$\PstarM{k}$}& \multicolumn{1}{c}{$\PstarM{(k+2)}$}& \multicolumn{1}{c}{$\SM{k}$}& \multicolumn{1}{c}{$\SM{(k+2)}$}& \multicolumn{1}{c}{$\SstarM{k}$}& \multicolumn{1}{c}{$\SstarM{(k+2)}$}\\ 
    \midrule
    \multirow{2}{*}{Acc} & RegExp1 & \textbf{1.00} & \textbf{1.00} & \textbf{1.00} & \textbf{1.00} & 0.91 & \textbf{1.00} & \textbf{1.00} & \textbf{1.00}\\
    & RegExp2 & \textbf{0.93} & 0.88 & 0.88 & \textbf{0.93} & 0.77 & 0.92 & 0.85 & 0.87\\
    \midrule
    \multirow{2}{*}{F1-score} & RegExp1 & \textbf{1.00} & \textbf{1.00} & \textbf{1.00} & \textbf{1.00} & 0.90 & \textbf{1.00} & \textbf{1.00} & \textbf{1.00}\\
    & RegExp2 & 0.93 & 0.86* & 0.87 & \textbf{0.95} & 0.71* & 0.92 & 0.81 & 0.87\\
    \bottomrule
    \end{tabular}
    }
\end{table}

In Table~\ref{tab:res-models-classifiers2}, we show the analysis of best accuracy and corresponding F1-score for the models vs. classifiers comparison. It can be observed that the results improved significantly as compared to the ones presented in Tab.~\ref{tab:res-models-classifiers}. We can also note that with this benchmark, only the $\SM{k}$ model failed to achieve perfect scores. Clearly, there are no significant differences between classifiers as they performed equally well regardless of the model.

\begin{table}[htb]
    \centering
    \caption{Best accuracy and corresponding F1-score metrics obtained by the NFAs for the analyzed classifiers}
    \label{tab:res-models-classifiers2}
    \resizebox{\columnwidth}{!}{
    \begin{tabular}{l|rrrr|rrrr}
    \toprule
    & \multicolumn{4}{c|}{Accuracy} & \multicolumn{4}{c}{F1-score}\\
    Model & \multicolumn{1}{c}{$\cmm$} & \multicolumn{1}{c}{$\cma$} & \multicolumn{1}{c}{$\csm$} & \multicolumn{1}{c|}{$\csa$} & \multicolumn{1}{c}{$\cmm$} & \multicolumn{1}{c}{$\cma$} & \multicolumn{1}{c}{$\csm$} & \multicolumn{1}{c}{$\csa$} \\
    \midrule
    $\PM{k}$ & \textbf{1.00} & \textbf{1.00} & \textbf{1.00} & \textbf{1.00} & \textbf{1.00} & \textbf{1.00} & \textbf{1.00} & \textbf{1.00}\\
    $\PM{(k+2)}$ & \textbf{1.00} & \textbf{1.00} & \textbf{1.00} & \textbf{1.00} & \textbf{1.00} & \textbf{1.00} & \textbf{1.00} & \textbf{1.00}\\
    $\PstarM{k}$ & \textbf{1.00} & \textbf{1.00} & \textbf{1.00} & \textbf{1.00} & \textbf{1.00} & \textbf{1.00} & \textbf{1.00} & \textbf{1.00}\\
    $\PstarM{(k+2)}$ & \textbf{1.00} & \textbf{1.00} & \textbf{1.00} & \textbf{1.00} & \textbf{1.00} & \textbf{1.00} & \textbf{1.00} & \textbf{1.00}\\
    $\SM{k}$ & 0.91 & 0.91 & 0.91 & 0.91 & 0.90 & 0.90 & 0.90 & 0.90\\
    $\SM{(k+2)}$ & \textbf{1.00} & \textbf{1.00} & \textbf{1.00} & \textbf{1.00} & \textbf{1.00} & \textbf{1.00} & \textbf{1.00} & \textbf{1.00}\\
    $\SstarM{k}$ & \textbf{1.00} & \textbf{1.00} & \textbf{1.00} & \textbf{1.00} & \textbf{1.00} & \textbf{1.00} & \textbf{1.00} & \textbf{1.00}\\
    $\SstarM{(k+2)}$ & \textbf{1.00} & \textbf{1.00} & \textbf{1.00} & \textbf{1.00} & \textbf{1.00} & \textbf{1.00} & \textbf{1.00} & \textbf{1.00}\\
    \bottomrule
    \end{tabular}
    }
\end{table} 



\section{Conclusions}
\label{sec:conclusion}

In this paper, we have proposed a method to transform an NFA with three types of states (accepting, rejecting and non-conclusive) to a weighted frequency automaton, which could be further transformed into a probabilistic NFA. The developed transformation process is generic since it allows to control the relative importance of the different types of states and/or transitions by customizable weights.

We have evaluated the proposed probabilistic automata on the classification task performed over two distinct benchmarks. The first one, based on real-life samples of peptide sequences proved to be quite challenging, yielding relatively low quality metrics. The second benchmark, based on a random sampling of a language described by a regular expression enabled us to show the power of probabilistic NFA, producing accuracy scores of 0.81--1.00 with F1-score ranging between 0.69 up to 1.00. The second benchmark allowed us to prove that given a representative sample of an underlying language, the probabilistic NFA can achieve very good classification quality, even without sophisticated parameter tuning.

In the future, we plan to apply some heuristics to tune the weights so that the classifiers perform even better, especially for real-life benchmarks. Given the generic nature of the proposed weighted-frequency automata we also plan to consider using a parallel ensemble of classifiers, differing not only in terms of weights, but also in how probabilities are combined.  


\bibliographystyle{splncs04}
\bibliography{main.bib}

\end{document}